\documentclass[10pt]{iopart}
\usepackage{epsfig}
\usepackage{supertabular}

\begin{document}
\title {The Network of Collaboration Among Rappers and its Community Structure}
\author {Reginald D. Smith}
\address{*Massachusetts Institute of Technology, Sloan School of
Management, Building E52, 77 Massachusetts Avenue, Cambridge, MA
02139-4307}
\address{Bouchet-Franklin Research Institute, P.O. Box 360262
,Decatur, GA 30036-0262}
\date {December 1,2005}
\ead {rsmith@sloan.mit.edu}


\begin {abstract}
The social network formed by the collaboration between rappers is
studied using standard statistical techniques for analyzing complex
networks. In addition, the community structure of the rap music
community is analyzed using a new method that uses weighted edges to
determine which connections are most important and revealing among
all the communities. The results of this method as well as possible
reasons for the structure of the rap music community are discussed.

\end {abstract}
\maketitle
Over the last decade, there has been a revolution in our
understanding of networks that permeate many aspects of our
universe. Among the many networks studied have been the
Internet\cite{Faloutsos}, metabolic pathways\cite{metabolic}, sexual
contacts\cite{sexual}, instant messaging\cite{instant message},
scientific collaborations\cite{scicollab1, scicollab2},
Congressional committees\cite{congress}, and even comic book
characters\cite{comics}. On a note related to this paper, the
network of shared online personal music libraries also has network
characteristics and can collectively define official (and
unofficial) music genres \cite{genre}. What all of these networks
exhibit is small world behavior, which is the behavior that the
average shortest path between any two nodes in a network is
extremely small compared to the size of the network. They also
exhibit scale-free characteristics in their degree distributions
that indicate power law scaling among the number of edge degree
among nodes.
The relevant characteristics and research regarding these networks
is well covered in several review articles\cite{review1, review2,
review3}. In this paper, the techniques used to analyze these
networks will be applied to collaboration among rap artists. In
addition, new methods of analyzing the community structure of
networks will be introduced using the rap network as an example.

\section {Hip-Hop Collaboration}

Rap as a music form is a subset of a larger cultural force formally
known as hip-hop. Rap artists are undoubtedly hip-hop's most visible
(and financially lucrative) manifestation, however, the hip-hop
community contains many other aspects including spoken word poetry,
turntables, break dancing, and graffiti art. One of the most
interesting aspects of hip-hop, particularly rap, is the amount of
collaboration between individuals. In rap music, different artists
belong to different record labels and groups like any other musical
genre. However, rap distinguishes itself from many other music forms
because there is frequent collaboration across group, music label,
or regional boundaries on specific songs. It is not unusual for two
rappers from different groups and labels to have cameo appearances
on the songs on each other's CDs. Though the writer would not argue
this makes rap superior to any other musical art forms, this aspect
does make it unique except for a few other genres such as jazz whose
collaboration network among early musicians is described in
\cite{jazzcite}.
The nature of rap music collaboration is useful, however, not just because it is a network, but that it has a relatively well-defined and transparent
community structure. Many networks have vague (or unknown to the researcher) community structures above the clique level. Rap music, however, has well
defined communities. Rap could be broken into several layers of organization like those shown below
\begin{enumerate}
\item Individual rappers
\item Groups/Supergroups (cliques)
\item Music Labels
\item Regional/Community affiliation
\end{enumerate}
The individual rappers are the nodes in the network studied here and
are self-explanatory. Groups and supergroups are common in rap.
Supergroups are groups of rappers who rap as a group but also
frequently release their own solo albums independent from the group
(but usually on the same music label). Music Labels are the
companies which contract the rappers and can contain hundreds of
artists. Often there is frequent collaboration among rappers in
music labels, however, it does not mean that the music labels are
cliques. Regional and community affiliation is probably the highest
level of community. It refers to the loose knit status of being part
of a ''region'' such as the Southern United States, the West Coast
US (mostly California),the New York City area, or even countries in
Europe. In addition, there are non-regional communities like
''underground'' rap which consists of rappers that are usually not
signed to major record labels and are not widely released
commercially.  In regions/communities there is a great deal of
collaboration, though much less than within a music label, that is
nevertheless more tightly knit than the overall rap community
(compare clustering coefficients in Tables \ref{rapstatistics} and
\ref{regions}).

\section{Methodology}

The main source for the data in this paper was the Internet website
the Original Hip-Hop Lyrics Archive (www.ohhla.com)\cite{ohhla}. The
hip-hop lyrics archive contains tens of thousands of fan submitted
lyrics for rap songs in several languages. In addition, it has a
standardized format on each song lyric where all artists in the song
are listed on the first line of the lyrics text file. This fact made
it easy to use a computer program to strip the names of rappers from
each song for analysis of the collaboration network once the network
was downloaded. Additional information was also provided by the huge
music database at AllMusic (www.allmusic.com)\cite{allmusic} and
from the rap news and information site AllHipHop
(www.allhiphop.com)\cite{allhiphop}.
Analysis was complicated, however, by the fact that the data was far
from "clean". Since individual fans submit the lyrics there are
often incorrect or inconsistent spellings of the names of rap
artists or groups. Many rap artists also have multiple pseudonyms.
This made any real analysis impossible without standardizing the
names. Unfortunately, this was an extremely tedious process. There
were several main techniques used to clean the data. One of the most
important was the use of a fuzzy search algorithm to match similarly
spelled artists. I used the Python programming language module
"agrepy" \cite{agrepy} by Michael Wise which is a Python port of the
popular UNIX fuzzy search algorithm 'agrep'. Using this algorithm I
was able to find similar misspellings. Using the search results,
correct spellings from AllMusic and Ohhla.com, and my own knowledge
I was able to correct inconsistent misspellings and standardize the
spellings of the vast majority of the rappers and groups.
Even after the data was standardized another problem arose that many
rappers, especially those in supergroups, frequently collaborated
solo with other rappers. This was an issue in accurately
representing the network since in one song they would be credited
only as the group and in another song they would be credited as an
individual. In order, to disambiguate the results I used all three
websites and personal knowledge to write down an extensive list of
rap groups. Then, using the data from the web sites I recorded in a
separate file the artist members of each group. In the final file, I
replaced all group names with the names of the individual artists. A
final issue is that some of the artists in the network are not
purely rappers. Many rappers have collaborated with other artists
from R\&B, funk, pop, rock, and other genres and they were also
included in the network since it would be difficult to disambiguate
them. I will argue these major data cleanings, and many more minor
ones, were extremely extensive though I cannot claim they are
completely comprehensive. The major players in the network are all
represented, however, some groups and artists had little data
available on them and were left as originally entered. It is my
contention, however, that the network reflects the actual rap
collaboration network very accurately. The song data analyzed in
this paper represents the network as of June 15, 2005 when the
source files were downloaded.

\section{Network Analysis}
\begin{table}[!htb]
\caption{Basic Networks Statistics and Comparable Undirected
Networks. $n$ is the number of nodes in the network, $M$ is the
number of undirected edges, $z$ is the average degree per
node,$\overline{\ell}$ is the average shortest path between any two
nodes, $C_1$ and $C_2$ are two measures for the clustering
coefficient. $r$ is the degree correlation coefficient and $\alpha$
is the power-law scaling exponent. All comparable numbers are from
\cite{review1} except the jazz musicians \cite{jazzcite} and
Brazilian popular music \cite{brazilpop}. Original papers for actors
are \cite{movie1,movie2} and company directors are
\cite{director1,director2}}
\begin{center}
\label{rapstatistics}
\begin{tabular}{|p{40pt}|c|c|c|c|c|c|c|c|c}
\hline
Network&$n$&$M$&$z$&$\overline{\ell}$&$C_1$&$\overline{C_2}$&$r$&$\alpha$ \\
\hline
Rappers&5533&57972&20.95&3.9&0.18&0.48&0.06&3.5\\
Movie Actors&449913&25516482&113.43&3.48&0.2&0.78&0.208&2.3\\
Board Directors&7673&55392&14.44&4.6&0.59&0.88&0.276&--\\
Jazz Musicians&1275&38326&60.3&2.79&0.33&0.89&0.05&-- \\
Brazilian Pop Music&5834&507005&173.8&2.3&--&0.84&--&2.57 \\
\hline
\end{tabular}
\end{center}
\end{table}

The rap collaboration network studied contained 6,552 rappers and
groups and over 30,000 songs that yielded 57,972 distinct edges. Of
these rappers, only 5,533 had at least one edge. The others had no
collaborations or were groups who were removed from the file and
replaced by their members. The rap collaboration network, like
almost all social networks, is considered to contain undirected
edges. The basic network analysis results are summarized in Table
\ref{rapstatistics}. First, and most importantly, the rap music
collaboration network exhibits small-world character with an average
shortest path $\overline{\ell}$ of 3.9. The network has a moderately
high average clustering coefficients\cite{review1}, $C_1$ which is
0.18 and calculated for the entire network using the equation
\begin{equation}
C_1 = \frac {\textrm {3 x number of triangles in the network}}
{\textrm{number of connected triples of vertices}}
\end{equation}
where a connected triple is a single vertex with edges running to an
unordered pair of others.
 The
$\overline{C}_2$,
 of 0.48 calculated by finding the $\overline{C_2}$ over all nodes where for
 each node $C_2$ is calculated by
 \begin{equation}
C_2 = \frac {\textrm{number of triangles connected to vertex i}}
{\textrm{number of triples centered on vertex i}}
 \end{equation}
 31.8\% of the nodes have a $C_2$ of 1. These two metrics identify rappers as members of a small-world community.
The scaling exponent, $\alpha$, is high at 3.5 and was calculated
using the equation for a scaling exponent from \cite{powerexponent}:
\begin{equation}
\alpha =
1+n\bigg[{\sum_{i=1}^n{\ln{\frac{x_i}{x_{min}}}}}\bigg]^{-1}
\label{scaleexponent}
\end{equation}
where $x_i$ is the degree of node $i$ and the $x_{min}$ is the
smallest node degree over which scaling behavior occurs. In this
paper $x_{min}$ used was 6.
\begin{figure}

    \centering
    \includegraphics[height=2in, width=2in]{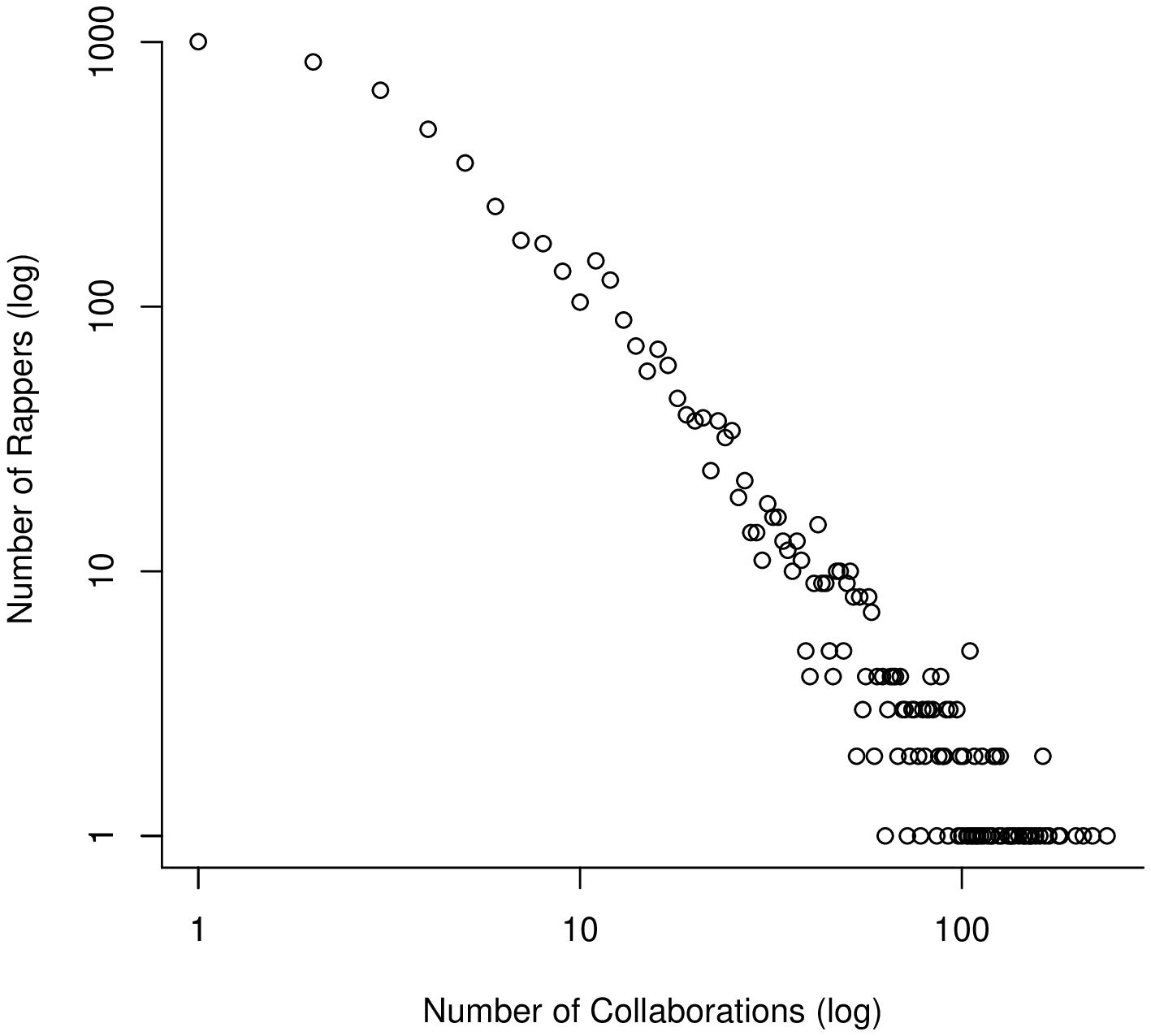}
    \includegraphics[height=2in, width=2in]{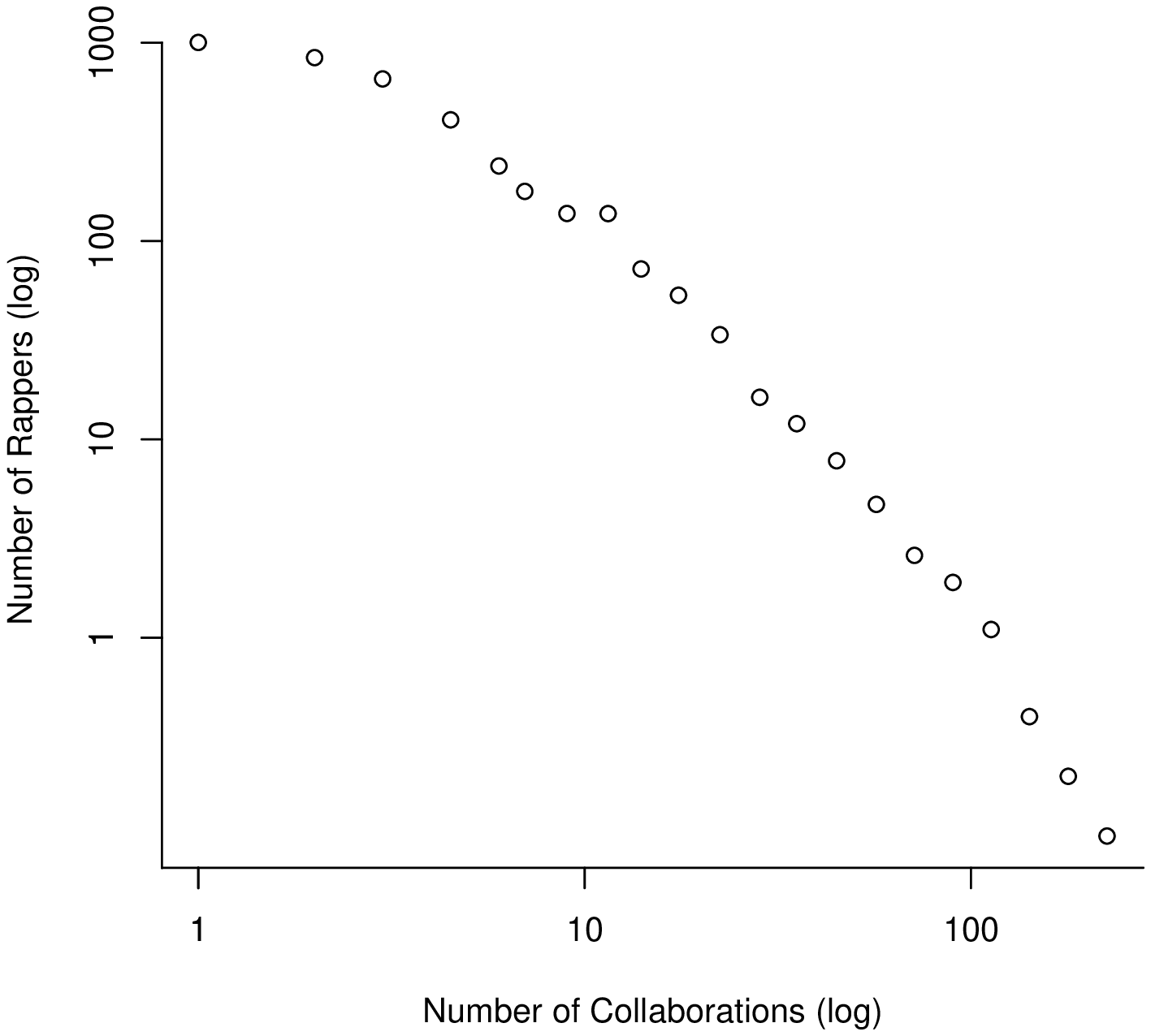}

        \caption{Log-log degree distribution of the number of rappers vs. the number of independent collaborators. The left plot is the plot of all data points. The right Log-log plot
        was created using log binning where the size of the bins grew exponentially (according to base 10). The
        average number of rappers was calculated for each bin and plotted at the geometric mean of the values in each bin.}

\end{figure}

\subsection{Assortative Mixing}

One of the most interesting results from basic network analysis is
the assortativity of the network as measured by the degree
correlation coefficient, $r$. Though $r$ is positive like almost all
other social networks, it has a very low value that would make the
network seem ambiguously assortative since $r$ has a value close to
zero.
A $r$ of near zero signifies that high degree nodes do not have a
disproportionately large preference for either high-degree or
low-degree node neighbors. This is in stark contrast to most social
networks which show a substantial amount of assortative mixing with
larger, positive values of $r$. Why do high degree nodes (both
well-connected and popular rappers) not have an affinity for each
other in the rap network unlike other social networks?

My first theory was the geographic regionalization of the rap
community. Rap music tends to be very regionalized and
collaborations often reflect this. The main regions of rap music in
the US can be divided into roughly five areas:

\begin{enumerate}
\item East Coast (most prominently New York and Philadelphia)
\item West Coast (California)
\item Southeast (Atlanta,  southern Florida, and adjacent areas, unfortunately often referred to as the "Dirty South")
\item Gulf Coast (most prominently New Orleans, Houston, and to a lesser extent, Memphis)
\item Midwest (spanning Chicago, St. Louis, Detroit, Cleveland, and other Midwest urban areas).
\end{enumerate}
When rappers collaborate outside of their group or record label, it
is usually with other rappers in their geographic area. There are
exceptions but collaboration is very regional. Therefore, I tried
measuring the degree correlation coefficient among the top 250
rappers as a whole as well as the coefficient among the regional
groupings of the rappers in this top 250. The results in Table
\ref{regions} demonstrate that apart from the Gulf Coast region,
none of the regions demonstrate assortative mixing among the top
regional players. Therefore, the lack of assortative mixing occurs
at many levels.
\begin{table}[!tb]
\caption{Regional Degree Correlation Coefficients in Four Regions.
Midwest not included because of small size (n=12)} \label{regions}
\begin{tabular}{|r|c|c|c|c|c|}
\cline{1-6} Region&$n$&$M$&$r$&$C_1$&$\overline{C_2}$ \\
\hline
Top 250&250&7438&0.037&0.31&0.38 \\
East Coast&135&3054&-0.01&0.37&0.43 \\
West Coast&47&690&-0.03&0.52&0.58 \\
Gulf Coast&36&386&0.211&0.70&0.70 \\
Southeast&20&130&-0.12&0.49&0.52\\
\hline
\end{tabular}
\end{table}
The reasons for no assortative mixing are likely many and complex.
First of all, unlike many social networks, rappers must be
understood in the context that they are often market players
competing with each other for album sales. Outside of their mutually
sustaining groups and record labels, there may actually be a
disincentive to the free collaboration among popular artists. One
does not want to help the record sales of his or her rival. Another
factor limiting assortative mixing is regional rivalries. This
especially reached its peak in the mid 1990s during the infamous
East Coast and West Coast feud centered between the rival record
labels Death Row Records, with artists like Snoop Dogg and 2Pac,
centered in Los Angeles and Bad Boy Records, with artists such as
Puff Daddy, Notorious B.I.G., and Ma\$e, centered in New York City.
This feud is believed to have helped fuel the climate that led to
the tragic deaths of 2Pac and Notorious B.I.G. Smaller feuds such as
this one can prevent many popular rappers from collaborating, even
if they could benefit from such collaboration. Second, many popular
rappers may feel it more advantageous to support less popular
rappers in order to help them obtain exposure (especially if they
are in the same label). New rap acts often feature prominent names
on their most popular singles and first albums in order to help
attract listeners unfamiliar with them or their style. Finally,
geographic distance often reflects different social networks and
venues which fuel collaboration. Though this effect cannot be
thought of as explanatory as demonstrated in Table \ref{regions}, it
is a limiting factor on rap collaborations. Whatever the factors,
they may be similar in effect to those affecting collaboration in
early jazz since the degree correlation coefficients of the rap and
jazz networks are so similar. In \cite{jazzcite} it was also shown
that assortative mixing was barely present in relationships between
jazz musicians. Racial segregation was mentioned as important factor
that shaped the network and may have played a part in limiting
assortativity similar to those postulated in rap.

\subsection{Most Connected Rappers}

A landmark paper by Newman \cite{scicollab1, scicollab2} studied
scientific co-authorship networks and attempted to tackle the
ambitious question of who is the most connected scientist. Following
part of his methodology, Tables \ref{betweentable} and
\ref{degreetable} attempts to summarize the Top 50 most connected
rappers based on two different metrics: first the betweenness metric
which measures the proportion of shortest paths in the network that
pass through a rapper and second the total degrees of each rapper
which represents the number of different collaborations. Both
methods show a large agreement in their results. First of all, Snoop
Dogg is claimed by both methods to be the most connected rapper.
This is likely since Snoop has collaborated with a huge number of
artists and based himself out of both Los Angeles and New Orleans
mixing with several prominent rap networks.
What causes a rapper to become well connected? In order understand this relationship better I looked at both album sales and the year of the
first major release by the artist to try to find correlations. The record sale index was calculated using data from the Recording Industry
Association of America (RIAA) website that has a database on the number of Gold and Platinum albums a rapper has won. A Gold album is a sale of
500,000 records. A Platinum album is a sale of 1 million records with double Platinum being 2 million, etc. The index equation was

\begin{equation}
I = 2\sum_i{f(P_i)} + G \label{musicindex}
\end{equation}

Where $G$ is the number of Gold albums an artist has won (if Gold is
the highest designation for an album) and $f(P)$ is the coefficient
of each Platinum album. For a single Platinum album the coefficient
is 1, for a double Platinum album it is 2, etc. The sum indicates
that the relevant value for the index is the sum of the coefficients
for all (multi)platinum records an artist has sold. So if an artist
has recorded 2 single Platinum and 1 triple Platinum albums
$\sum_i{f(P_i)}$ is 5. Unfortunately, there was zero correlation
between either the betwenness score or the node degree of a rapper
and the record sales index. There was also no discernable
correlation with the starting year for a rapper and the two metrics.
Therefore, the connected status of a rapper, like assortativity,
likely does not have a single simple explanation. The variables
influencing how connected a rapper is can include perceptions of
talent, social stature and reputation, and even personal preference.
For an example, Dr. Dre and Snoop Dogg are both prominent West Coast
rappers. Dr. Dre only has a node degree of 105 compared to Snoop's
240 despite having a higher record index score and the same regional
roots. This is reflected for several reasons including Dr. Dre has
gone more into producing artists like Eminem than rapping and
typically has not collaborated as prolifically as Snoop over the
years.

The complex nature of collaboration may also explain the high
$\alpha$ of 3.5. Since only a few versatile rappers have a high
degree, the degree distribution drops off more sharply than in
networks with less complex forms of preferential attachment. One
interesting note, however, is though the year a rapper began rapping
does not determine their rank, there is a clear trend ($R^2$ = 0.55
t-stat = 7.7) that the average number of new collaborators per year,
calculated by dividing the degree by the number of years since the
debut, is steadily increasing. So it seems newer rappers are more
apt to collaborate than older ones. Perhaps this is connected to the
increasing commercial prominence and rapid growth of rap over the
late 1990s.

\subsection{Community Structure}

Early in the paper, a rough outline of the community hierarchical
structure of rap music was given. Though this rap network can be
readily apparent to rap fans or critics, it can be difficult to
extract community boundaries using automatic algorithms. Many graph
finding algorithms such as the fast modularity community structure
algorithm\cite{fastcom} and the clique percolation
method\cite{cliqueperc} tend to either only correctly assign groups
(cliques) or overestimate the size of larger groupings (groups or
geographical regions). The clique percolation method identifies
groups and sometimes identifies geographical regions correctly but
has trouble focusing on identifying music labels. The fast
modularity community structure algorithm only can recognize some of
the small and peripheral rap groups.
In order to clarify different levels of group structure it can be
advantageous to not just analyze the topology of the network but the
types of interactions among its participants. In particular, the
data allowed not only the identification of edges in the network,
but also the frequency which a certain edge (collaboration)
occurred.

In order to take advantage of the frequencies of collaboration, they
can be used to accentuate the differences between frequent
collaboration partners and more casual ones. In particular the
following algorithm was used to generate a new network from the
data:
\begin{enumerate}
\item Create a weighted adjacency matrix of edges where the weight of an edge is the number of times a given collaboration occurs in the data set.
\item Determine the value of the highest weighted edge for every node
\item Use the following equation to ''mark'' edges corresponding to each node if they don't meet the following criteria

[Edge Weight]$^2$  $\geq$  X\%[Max Edge Weight for Node]$^2$

\item If an edge is "marked" by both of the nodes it is connected to, remove the edge from the network (set both entries in the adjacency matrix to 0).
\item Generate the new edge list for the network
\end{enumerate}

\begin{figure}[!ptb]
\begin{center}
\includegraphics[width=\textwidth]{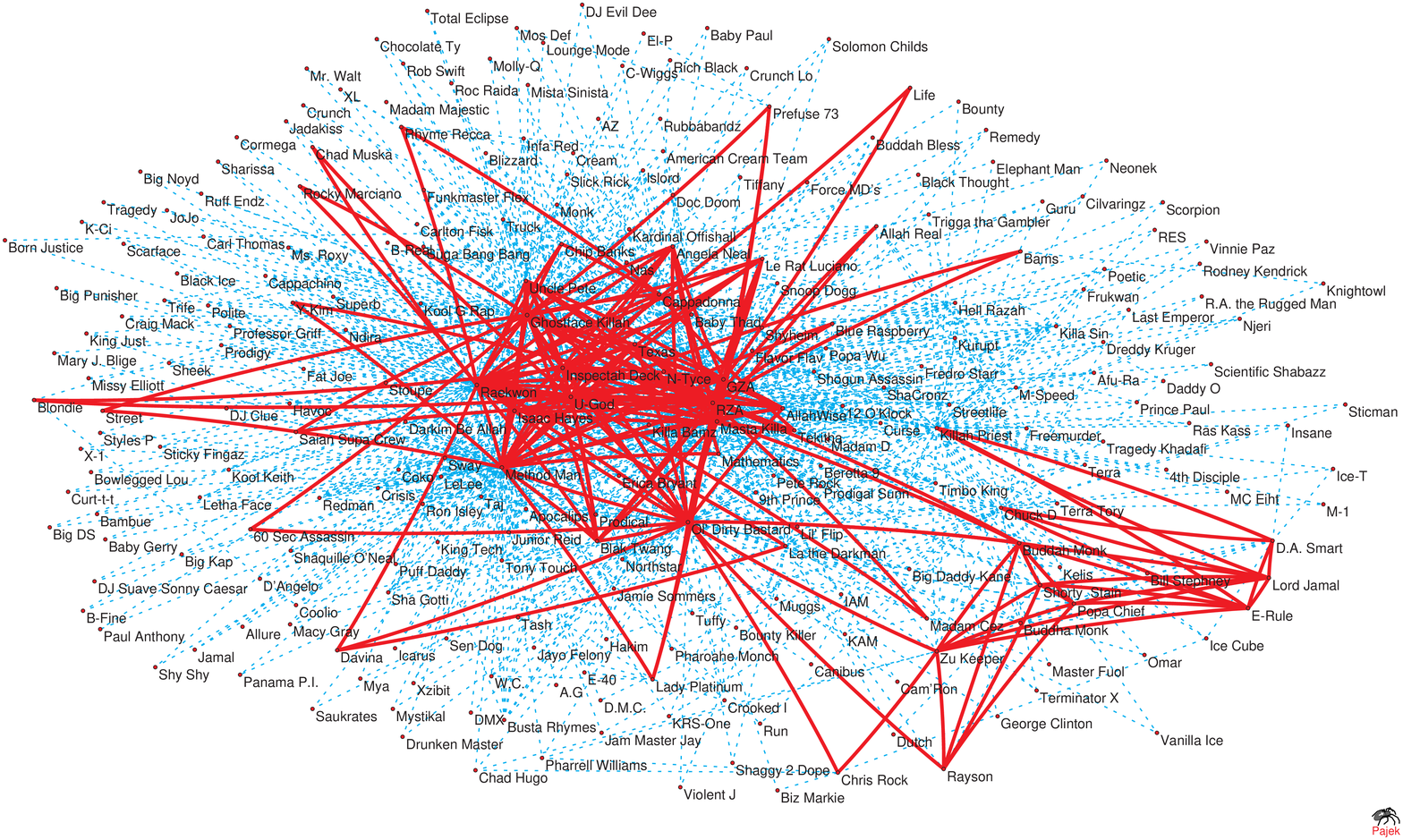}
\end{center}
\caption{A collection of adjacent k=3 k-cliques centering on the
rapper RZA found using the clique percolation method after the
weighted edge disparity algorithm is run for X = 50. The community
has red edges and sits over the network of all neighbors of the
nodes in the community. All of the rappers with several exceptions
such as Chuck D, Isaac Hayes, and Chris Rock are directly or
indirectly affiliated with the Wu-Tang supergroup and their music
labels. The highly clustered rappers in the middle of the diagram
are the core members of the original Wu-Tang Clan group (GZA, RZA,
Ol' Dirty Bastard, Method Man, Raekwon, Ghostface Killah, Inspectah
Deck, Masta Killa, U-God). Plotted with Kamada-Kawai graphing
algorithm.} \label{networks}
\end{figure}

In the previous equation, $X$ represents a percentage from 0-100
chosen in order to extract certain given features of the network.
Step 3 uses a squaring of the edge weights in order to create a
large enough disparity (especially if edges on a node have low
weights) that allows us to extract the most important edges. Step 4
is to assure we do not generate a directed edge since it is possible
for an edge to be marked by one of its nodes but the other.
The new network created is accentuated by only the most important
connections for each node being retained which throws much of the
community structure in sharper relief. Around $X$=10 most of the
music label and community affiliation is visible. At $X$=50 and
higher there is a clear separation of music labels and communities
from larger geographical affiliation. The red edges in Figure
\ref{networks} show the results of applying the edge disparity
algorithm at X=50 and then applying the clique percolation method,
for the union of adjacent k-cliques where k=3 and centered at the
RZA, a member of the supergroup known as the Wu-Tang Clan. After the
two algorithms are applied, the network is limited almost to only to
the Wu-Tang Clan and its affiliate rappers. In Figures 3-6, the
Wu-Tang clan and its affiliate rappers are shown, but so are their
hundreds of neighbors. The differences in the network for X=10, 50,
\& 90 is shown to demonstrate how the edge disparity algorithm
accentuates the most important relationships. The relationships in
Figure \ref{networks} also reflect deeper relationships in the
Wu-Tang community. For example, Raekwon ''discovered'' LA the
Darkman and collaborated most with him and the reduced network shows
this relationship extremely clearly. The fast modularity community
algorithm also has greater success with the refined network
separating not only smaller rap affiliations but several major ones
as well as some underground and Christian rap communities. It should
be noted that the edge disparity algorithm alone does not find
communities but should be used as a tool to refine networks with
weighted collaborations for analysis with other community
identification algorithms.

\section{Discussion}

The network of collaboration among rappers in songs is a small world
network which follows different rules of organization than typical
social networks. One of the current questions regarding the nature
of networks is the origin of assortative or disassortative mixing in
networks. The preponderance of evidence points to social networks as
largely being assortative while natural networks being large
disassortative. Whether there is an inherent sociological mechanism
that causes this disparity is still a matter of debate, the rap
network allows us to recognize that under some constraints or
organization, the assortative mixing aspect of social networks can
be more muted. Given the regional and affiliate nature of rap
collaboration, perhaps it is better to interpret the rap
collaboration network as a union of smaller networks based off of
other types of affiliations that are not readily apparent. The rap
collaboration network may also be the ''shadow'' of another network
which includes rappers, producers, turntablists, and other members
of the music scene whose interaction forms a more traditional social
network and is incompletely recognized by solely looking at rap
collaborations in commercial albums. In fact, many informal and
non-commercial song collaborations are not covered by the song
lyrics database at ohhla.com.

The community structure of rap also supports the assertion of many
\cite{scicollab2,weighted1,weighted2,weighted3} that community
structure in networks should use weighted edges instead of just a
binary edge topology. Many algorithms are designed to search out
community structure in networks based only on topological criteria
giving equal weight to all edges. Although this can provide much
insight into network structure, it is likely that community
structure is not most clearly defined by relationships alone. Even
when factors are taken into account to remove edges from the
network, they are often based on the relationship between that edge
and some topological criterion. At the single node level, all edges
are assumed to be just as important. Rap is another example that
shows this may not always be the case. Heavily used associations can
give more illumination to the characteristics of communities than
rarely uses associations. In rap music, the level of collaboration
between two artists can help elucidate their actual community
connections. Similar factors may elucidate community connections in
other social networks. For example, in electronic communication
networks such as email and instant messenger, perhaps taking into
account the number of emails/messages or the byte size of data
exchanged among an edge over a fixed period can help show which
relationships are more important and which are more trivial. By
using weighted edges and interaction dynamics, as well as,
topological considerations, it is likely that the nature and
structure of communities will become much clearer.

\clearpage

\begin{table}
\caption{Top 50 Most Connected Rappers by Betweenness and
Corresponding Sales Index}
\label{betweentable}
\begin{tabular} {|r|c|c|}
\cline{1-3} Artist&Betweenness($*10^6$)&Sales Index \\

\hline
\hline
\hline
Snoop Dogg&29962&27 \\
Kurupt&28634&5 \\
2Pac&21993&41 \\
Busta Rhymes&21302&12 \\
Guru&19670&2 \\
Lil' Flip&16496&4 \\
Fat Joe&16318&3 \\
Method Man&16273&20 \\
Master P&15567&27 \\
Ol' Dirty Bastard&15036&15 \\
KRS-One&14680&2 \\
RZA&14604&14 \\
Redman&14373&9 \\
Jay-Z&14339&35 \\
Ghostface Killah&14134&16 \\
Nas&14122&13 \\
Xzibit&13236&4 \\
Scarface&12720&13 \\
Twista&12407&3 \\
Yukmouth&12104&2 \\
Killah Priest&12089&0 \\
E-40&12039&5 \\
Funkmaster Flex&11637&4 \\
Talib Kweli&11201&0 \\
Too \$hort&10584&21 \\
Daz Dillinger&10544&3 \\
Raekwon&10380&15 \\
Z-Ro&10293&0 \\
Missy Elliott&10286&12 \\
Havoc&10025&5 \\
Prodigy&9790&6 \\
Ice-T&9687&5 \\
Wyclef&9467&18 \\
Kool G Rap&9370&0 \\
Puff Daddy&9253&15 \\
Eminem&9234&50 \\
Nelly&9158&42 \\
Jermaine Dupri&9145&2 \\
Ice Cube&8957&31 \\
Warren G&8784&8 \\
Kool Keith&8525&0 \\
Ras Kass&8480&0 \\
Q-Tip&8461&1 \\
Big Daddy Kane&8332&2 \\
Juvenile&7859&15 \\
Bun B&7859&0 \\
Brotha Lynch Hung&7653&0 \\
MC Eiht&7563&1 \\
DMX&7473&23 \\
Lil' Kim&7306&8 \\
\hline

\end{tabular}
\end{table}

\begin{table}[!p]
\label{degreetable}
\caption{Top 50 Most Connected Rappers by Degree and Corresponding Sales Index and First Album Year}
\begin{tabular}{|r|c|c|c|}
\cline{1-4} Artist&Degree&Sales Index&First Album Year \\
\hline

Snoop Dogg&240&27&1994 \\
Busta Rhymes&220&12&1996 \\
Method Man&208&20&1994 \\
Kurupt&199&5&1995 \\
2Pac&181&41&1993 \\
Redman&179&9&1993 \\
Ol' Dirty Bastard&169&15&1994 \\
Master P&165&27&1996 \\
Funkmaster Flex&163&4&1997 \\
Nas&163&13&1996 \\
Jay-Z&160&35&1996 \\
Ghostface Killah&156&16&1994 \\
Raekwon&152&15&1994 \\
Fat Joe&151&3&1998 \\
Guru&149&2&1998 \\
Puff Daddy&146&15&1997 \\
RZA&145&14&1994 \\
Missy Elliott&141&12&1997 \\
Jermaine Dupri&137&2&1998 \\
Twista&135&3&1999 \\
Prodigy&134&6&1993 \\
KRS-One&132&2&1997 \\
Xzibit&127&4&2001 \\
E-40&126&5&1995 \\
Scarface&126&13&1993 \\
Havoc&125&5&1993 \\
Lil' Kim&123&8&1997 \\
Yukmouth&123&2&1995 \\
Daz Dillinger&121&3&1995 \\
Too \$hort&121&21&1989 \\
Killah Priest&120&0&1998 \\
Silkk the Shocker&119&8&1997 \\
DMX&117&23&1998 \\
Eminem&114&50&1999 \\
Common&113&1&1992 \\
Jadakiss&113&2&2001 \\
Noreaga&112&2&1998 \\
Foxy Brown&110&5&1996 \\
Q-Tip&109&1&1999 \\
Big Punisher&108&3&1998 \\
Nate Dogg&108&0&1997 \\
Wyclef&107&18&1996 \\
Ludacris&106&18&2000 \\
Ice Cube&105&31&1989 \\
Kool G Rap&105&0&1995 \\
Lil' Flip&105&4&2002 \\
Lil' Jon&105&9&1997 \\
Talib Kweli&105&0&1998 \\
Ma\$e&104&10&1997 \\
Z-Ro&103&0&1998 \\
\hline
\end{tabular}
\end{table}

\clearpage
\clearpage
\begin{figure}

    \centering
    \includegraphics[height=2in,width=\textwidth]{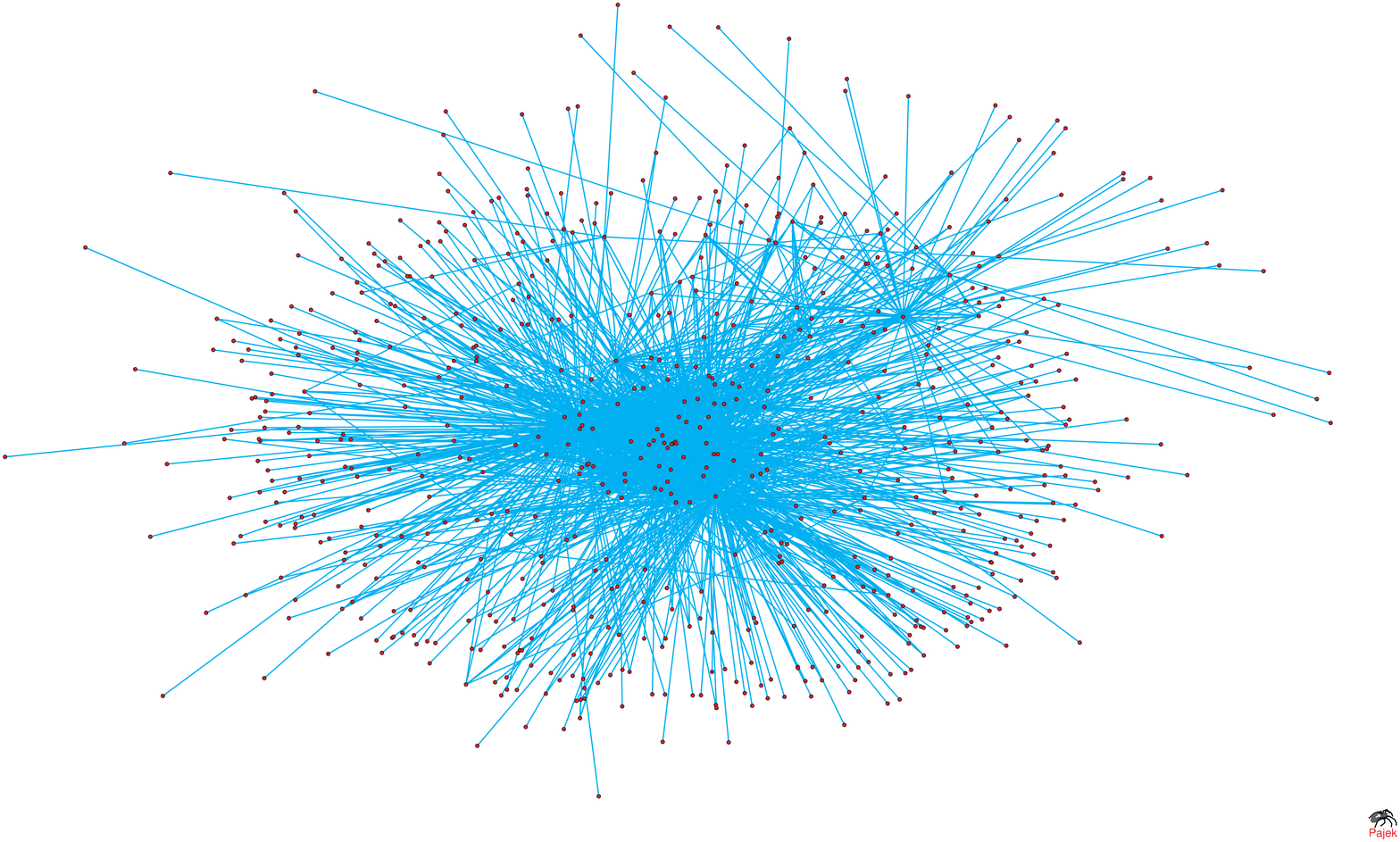}
    \caption{Wu Tang Clan and their neighbors. Plotted with Kamada-Kawai graphing algorithm.}
    \label{networks2}
    \includegraphics[height=2in,width=\textwidth]{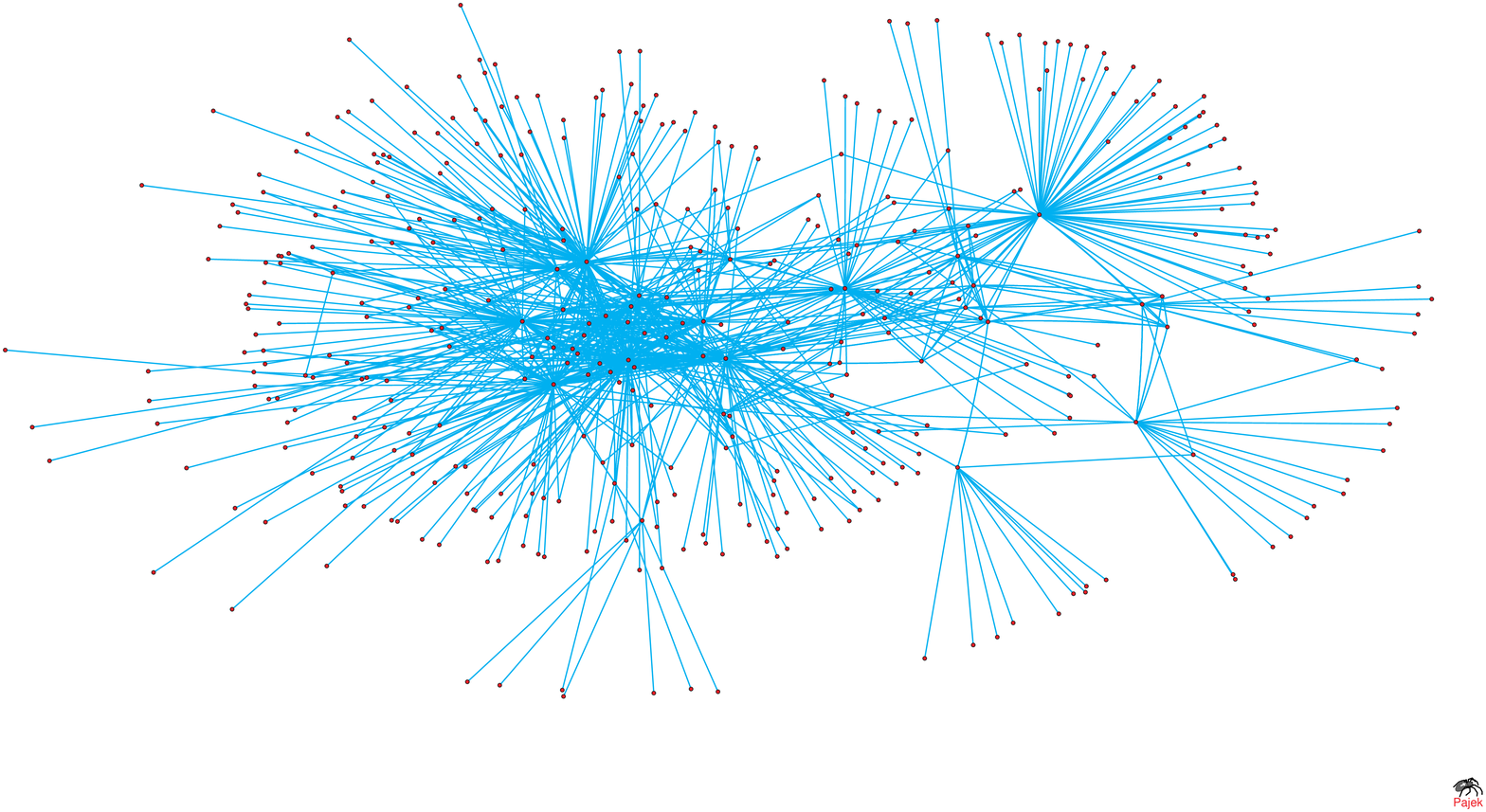}
    \caption{Wu Tang Clan and neighbors with edge disparity
    algorithm applied for X=10. Plotted with Kamada-Kawai graphing algorithm.}
    \label{networks3}
    \includegraphics[height=2in,width=\textwidth]{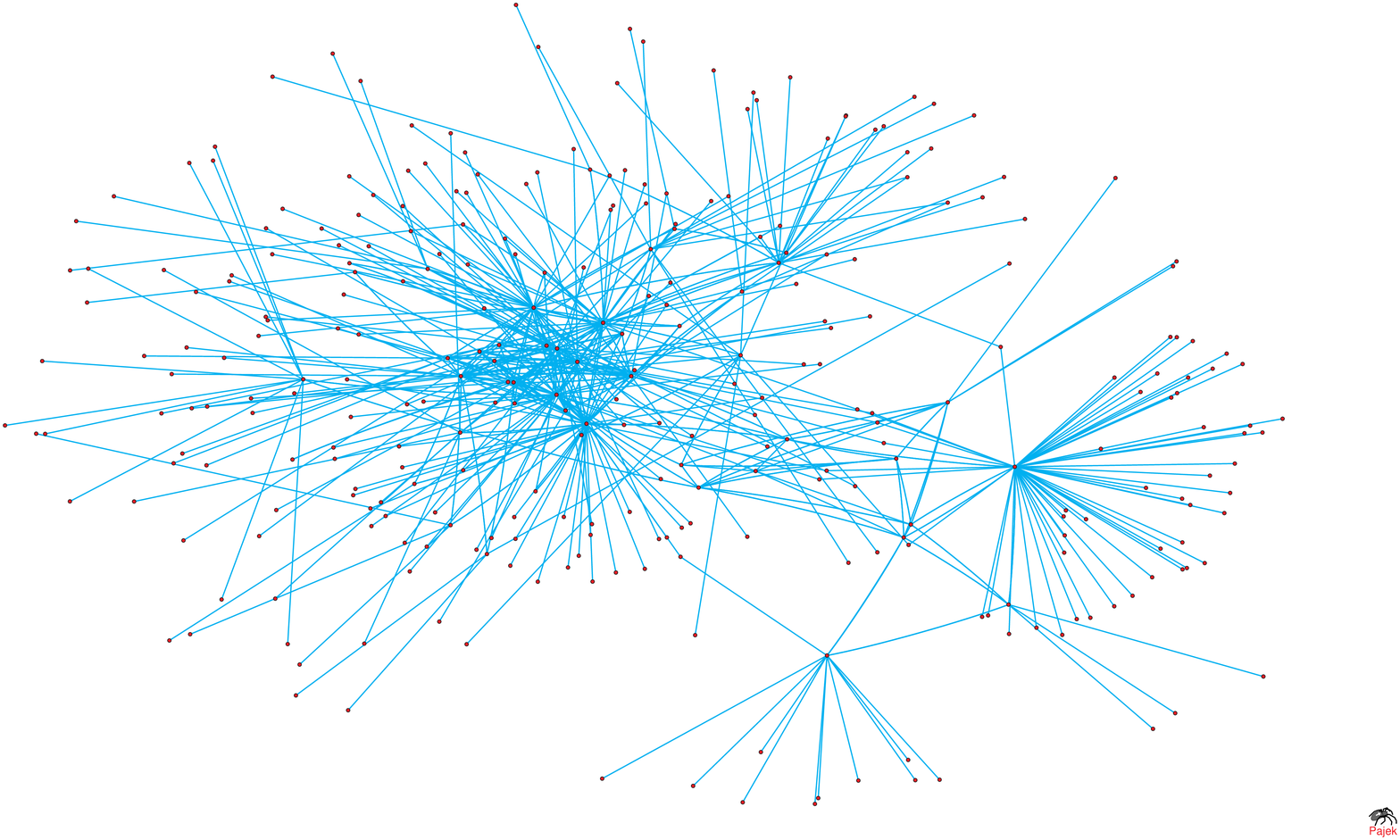}
        \caption{Wu Tang Clan and neighbors with edge disparity
    algorithm applied for X=50. Plotted with Kamada-Kawai graphing algorithm.}
    \label{networks4}
    \includegraphics[height=2in,width=\textwidth]{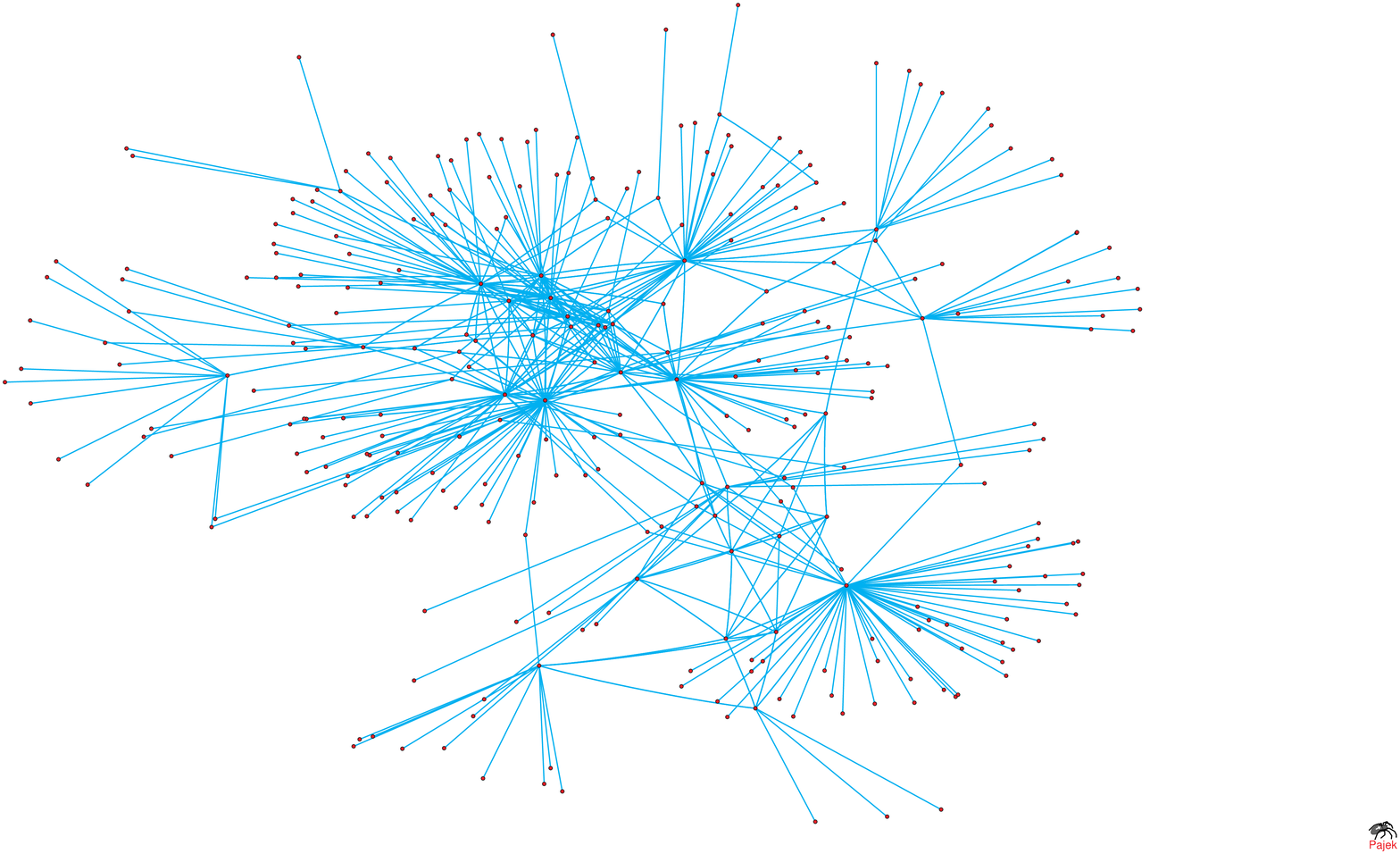}
        \caption{Wu Tang Clan and neighbors with edge disparity
    algorithm applied for X=90. Plotted with Kamada-Kawai graphing algorithm.}

\label{networks5}
\end{figure}

\clearpage

\section{Acknowledgements}
The author would like to thank Mark Newman for his helpful time and
comments regarding this paper. The author would also like to thank
Steve ''Flash'' Juon, the webmaster of the Original Hip-Hop Lyrics
Archive, for his advice on the nature of rap music collaboration as
well as Cameron Wadley for opinions regarding the accuracy of the
rapper connectedness rankings. The author would also like to
acknowledge the use of Pajek for network analysis, Agrepy for fuzzy
logic searching to clean up the data, and the programs for the fast
modularity community finding algorithm and CFinder for community
analysis.

\end{document}